\documentclass[twocolumn]{aastex631}

\shorttitle{NIR POLARIZATION CHARACTERISTICS OF THE ZL OBSERVED WITH DIRBE/COBE}
\shortauthors{Takimoto et al.}
\graphicspath{{./}{figures/}}

\begin{document}

\title{NEAR-INFRARED POLARIZATION CHARACTERISTICS OF THE ZODIACAL LIGHT OBSERVED WITH DIRBE/COBE}

\author[0000-0002-8405-9549]{Kohji Takimoto}
\affiliation{Department of Space Systems Engineering, School of Engineering, Kyushu Institute of Technology, Fukuoka 804-8550, Japan}

\author[0000-0002-5698-9634]{Shuji Matsuura}
\affiliation{Department of Physics and Astronomy, School of Science, Kwansei Gakuin University, Hyogo 669-1337, Japan}

\author[0000-0002-6468-8532]{Kei Sano}
\affiliation{Department of Space Systems Engineering, School of Engineering, Kyushu Institute of Technology, Fukuoka 804-8550, Japan}

\author[0000-0002-9330-8738]{Richard M. Feder}
\affiliation{Department of Physics, California Institute of Technology, Pasadena, CA 91125, USA}

\begin{abstract}

We report near-infrared polarization of the zodiacal light (ZL) measured from space by the Diffuse Infrared Background Experiment (DIRBE) onboard the Cosmic Background Explorer in photometric bands centered at 1.25, 2.2, and 3.5 $\mu$m.
To constrain the physical properties of interplanetary dust (IPD), we use DIRBE Weekly Sky Maps to investigate the solar elongation ($\epsilon$), ecliptic latitude ($\beta$), and wavelength ($\lambda$) dependence of ZL polarization.
We find that the polarization of the ZL varies as a function of $\epsilon$ and $\beta$, consistent with observed polarization at $\lambda$ = 550 nm. 
While the polarization dependence with wavelength at $(\epsilon$, $\beta)=(90^{\circ}$, $0^{\circ})$ is modest (increasing from 17.7 $\pm$ 0.2\% at 1.25 $\mu$m to 21.0 $\pm$ 0.3\% at 3.5 $\mu$m), the variation is more pronounced at the North Ecliptic Pole (23.1 $\pm$ 1.6, 35.1 $\pm$ 2.0 and 39.3 $\pm$ 2.1\% at 1.25, 2.2 and 3.5 $\mu$m, respectively).  
The variation of ZL polarization with wavelength is not explained by either Rayleigh scattering or by absorptive particles larger than 10 $\mu$m.
\end{abstract}

\section{Introduction} \label{sec:intro}
\begin{figure}[htbp]
\epsscale{1.2}
\plotone{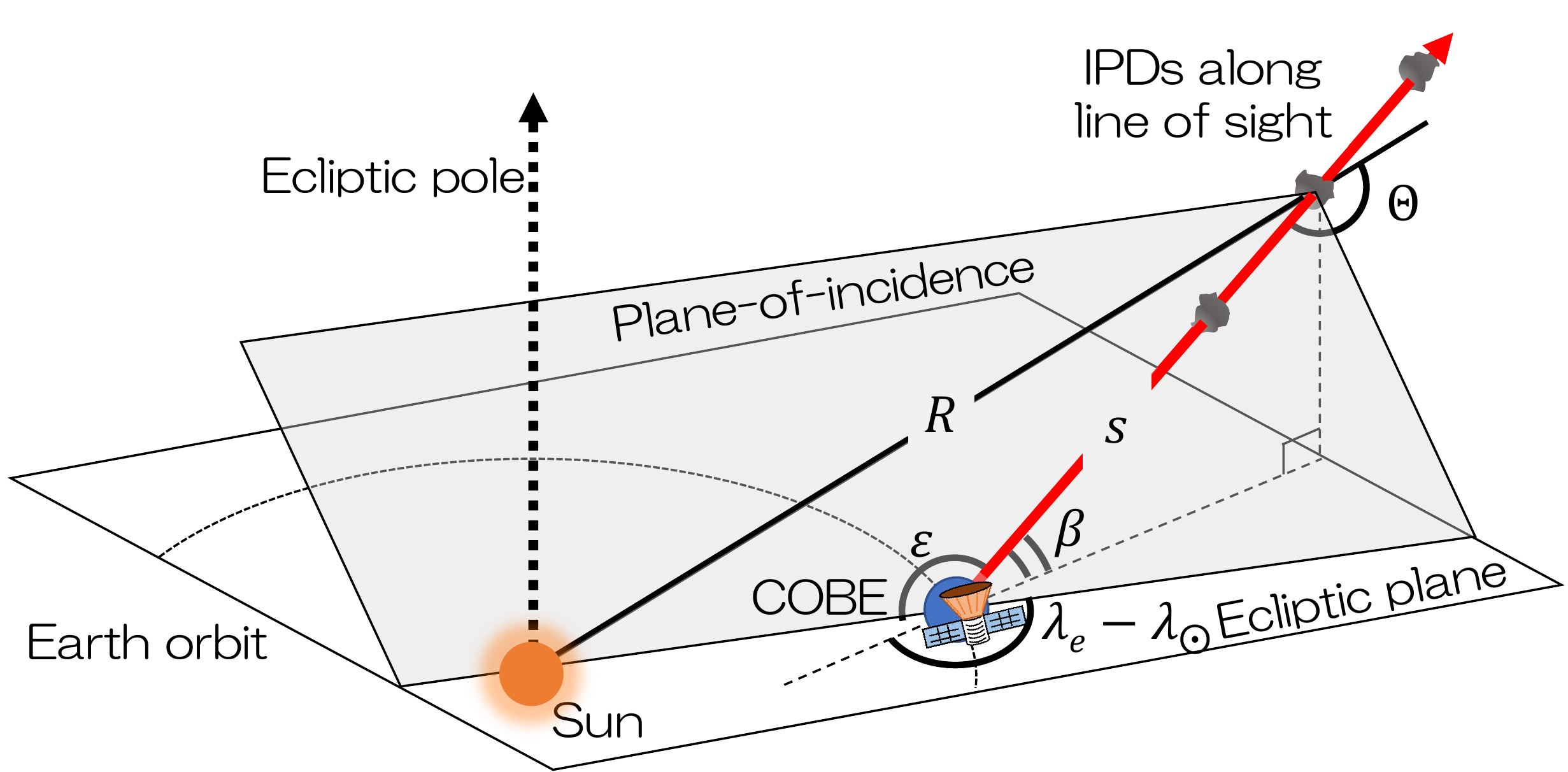}
\caption{Geometric description of COBE observation with respect to the ecliptic latitude $\beta$, and the helio-ecliptic longitude, ($\lambda_{e}-\lambda_{\odot}$), where $\lambda_{e}$ is the ecliptic longitude and $\lambda_{\odot}$ is the ecliptic longitude of the Sun. 
The red arrow represents the line of sight from DIRBE, and $s$ indicates the distance between DIRBE and a IPD grain. 
The observed ZL intensity corresponds to integrated light scattered by all IPD grains along DIRBE line of sight.
The radial distance from the grain to the Sun is denoted by $R$, and the solar elongation is indicated by $\epsilon$.
The scattering angle is denoted by $\Theta$.
}
\label{fig:ZLgeometry}
\end{figure}

The zodiacal light (ZL) in the optical and the near-infrared is caused by sunlight scattered from the interplanetary dust (IPD) particles in the solar system.
Most recent studies support that the IPD is primarily derived from comets \citep{1995P&SS...43..717L,2002Sci...296.1087S,2006AJ....132.1354F,2010ApJ...713..816N,Nesvorn__2011,2015ApJ...813...87Y} and asteroids \citep{1984Natur.312..505D,1989Metic..24...99S,1996PASJ...48L..47M,2010ApJ...719..394T}, however further investigation is required to determine the relative contribution of the two.
The typical radius of IPD particles is sub-micron to 100 $\mu$m \citep{1985Icar...62..244G,reach1988zodiacal}.
Measurements of the ZL intensity and polarization provide important information that can help constrain the composition of the IPD.
The Diffuse Infrared Background Experiment (DIRBE) onboard the Cosmic Background Explorer (COBE) characterizes the ZL in the near-infrared \citep{1991AIPC..222..161H,1992ApJ...397..420B,1993SPIE.2019..180S}.
\citet{1998ApJ...508...44K} modeled the scattered light component of the ZL intensity $I_{\lambda}$ at wavelength $\lambda$ as
\begin{equation}
I_{\lambda} = \int{n(X, Y, Z)A_{\lambda}F^{\odot}_{\lambda}\Phi_{\lambda}(\Theta)}ds,
\label{eq:kel}
\end{equation}
where $n(X, Y, Z)$ is the three-dimensional density distribution of the IPD component, $A_{\lambda}$ is the albedo, $F^{\odot}_{\lambda}$ is the solar flux, and $\Phi_{\lambda}(\Theta)$ is the phase function at scattering angle $\Theta$.
The ZL intensity observed in a given direction corresponds to the intensity integrated along the line of sight $s$.
We show the ZL geometry observed in helio-ecliptic coordinates in Figure \ref{fig:ZLgeometry}.

ZL scattered by IPD particles also exhibits systematic linear polarization.
ZL polarization measurements are important for constraining IPD parameters such as size, shape, and constituent minerals, as well as for photometric and spectroscopic observations.
The linear polarization $P$ is defined as the difference in the degree of polarization intensity along the plane $I_{\perp}$ perpendicular to the scattering plane and $I_{\parallel}$ parallel to the scattering plane:
$$
P=\frac{I_{\perp}-I_{\parallel}}{I_{\perp}+I_{\parallel}}.
$$
$P$ is positive if the electric field polarization direction is perpendicular to the scattering plane (Figure \ref{fig:ZLgeometry}).
Another source of near-infrared diffuse emission on a large scale is the Galactic plane, whose emission is due to stellar light and is polarized by less than 1\% due to absorption by magnetically aligned dust particles \citep{1990ApJ...348L..13N,1990AJ....100..274J,1990ApJ...357..113M}.

\begin{figure}[tbp]
\epsscale{1.2}
\plotone{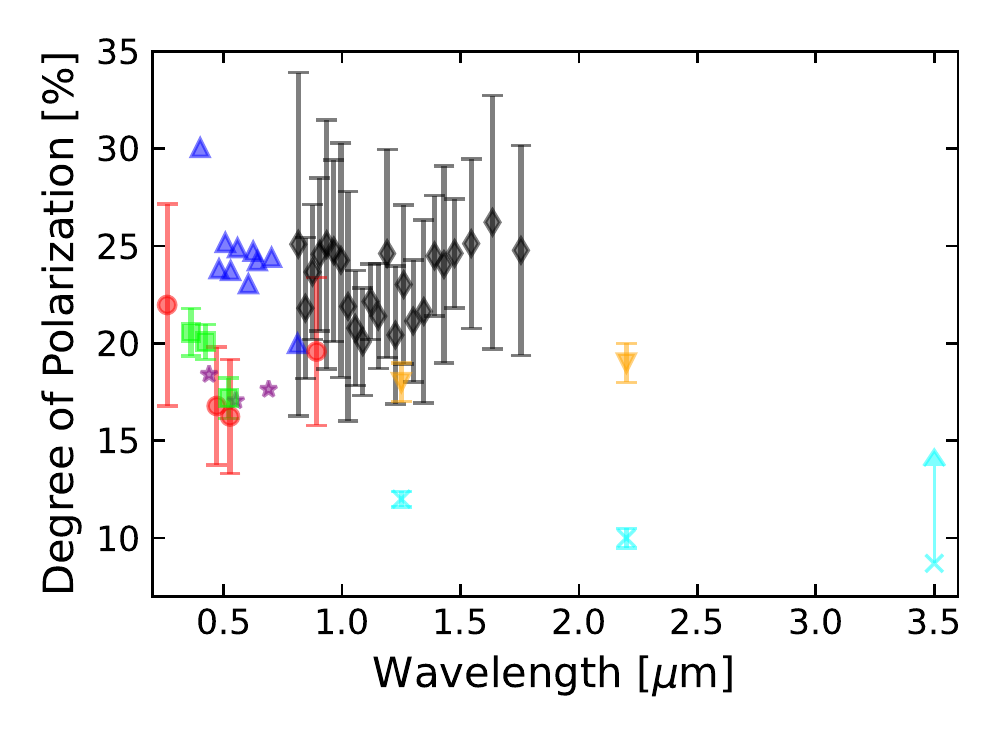}
\caption{Zodiacal light polarization observed at different positions.
Red circles: Astro 7 at $\epsilon$ = 30$^{\circ}$ \citep{1979A&A....74...15P}; blue triangles: Skylab at the North Celestial Pole \citep{1980IAUS...90...19W}; green squares: Helios at $\beta$ = 16$^{\circ}$, $\epsilon$ = 90$^{\circ}$ \citep{1982A&A...105..364L}; purple stars: an average of three similar results (OSO-5 at $\epsilon$ = 90$^{\circ}$, \citet{1972ApJ...174..705S}; balloon at $\epsilon$ = 30$^{\circ}$, \citet{1970ApJ...161..309V}, ground-based at $\epsilon$ = 39$^{\circ}$, \citet{1967NASSP.150...57W}); black diamonds: CIBER at the North Ecliptic Pole \citep{Takimoto_2022}; aqua crosses and orange inverted triangles: DIRBE at $\beta$ = 0$^{\circ}$, $\epsilon$ = 90$^{\circ}$ \citep{1994ApJ...431L..63B,Takimoto_2022}.
}
\label{fig:ZLpol_compare}
\end{figure}

Most measurements are in the optical as opposed to the infrared, because the near-infrared measurements are difficult to do from the ground due to statistical and systematic errors from strong airglow by OH radicals in the upper atmosphere which is several hundred times brighter than ZL \citep{1998A&AS..127....1L}.
Several studies have been reported on ZL polarization measurements in the visible band.
Figure \ref{fig:ZLpol_compare} shows the results of recent ZL polarization observations from the ground and from space.
Each observation was obtained in a different region of the sky.
ZL polarization at optical wavelengths is observed by Astro 7 \citep{1979A&A....74...15P}, Skylab \citep{1980IAUS...90...19W}, Helios \citep{1982A&A...105..364L}, OSO-5 \citep{1972ApJ...174..705S}, balloon borne experiment \citep{1970ApJ...161..309V}, and ground-based telescopes \citep{1967NASSP.150...57W}, and with no strong trend in degree of polarization from $\lambda$ = 0.45 $\mu$m to 0.8 $\mu$m.
\citet{2020P&SS..19004973L} reported a ZL polarization map at 550 nm based on data from \citet{1996ASPC..104..301L} and \citet{1998A&AS..127....1L}.

More recently, ZL polarization has been measured in the near-infrared.
The ZL polarization spectrum from $\lambda$ = 0.8 $\mu$m to 1.8 $\mu$m observed by CIBER \citep{Takimoto_2022} also shows little wavelength dependence.
Note that since these observations probe different regions, the origin of IPD is considered to be different.
In the case where wavelength dependence is observed for ZL polarization at wavelengths even longer than 1.8 $\mu$m, the dominant scattering property of IPD has changed from geometrical-optical scattering to Mie scattering.
Thus, it is possible that polarization observations in the near-infrared can constrain the particle radius of the dominant IPD causing ZL polarization.

DIRBE has performed the first and unique polarization survey of the diffuse sky at $\lambda$ = 1.25, 2.2, and 3.5 $\mu$m.
\citet{1994ApJ...431L..63B} reported the ZL polarization in the ecliptic plane (ecliptic latitude $\beta=0^{\circ}$) at solar elongation $\epsilon=90^{\circ}$ using only one week of observations, but diffuse radiation and starlight were not subtracted in all wavelength bands, and the 3.5 $\mu$m results include a thermal radiation component of the ZL.
\citet{Takimoto_2022} also estimated the ZL polarization at 1.25 and 2.2 $\mu$m from the DIRBE result presented in \citet{1994ApJ...431L..63B}, but ZL polarization shows little $\lambda$-dependence after considering other diffuse light sources \citep{1998ApJ...508...74A,2001ApJ...555..563C,2005A&A...436..895G,2006AJ....131.1163S,10.1093/pasj/65.6.120,2015ApJ...806...69A}.

In this paper, we report a new $\epsilon$-, $\beta$-, and $\lambda$-dependence of the ZL polarization in the near-infrared measured from space by DIRBE.
We obtain the ZL polarization using 41 weeks of DIRBE observations and make near-infrared ZL polarization maps for the three bands from 1.25 $\mu$m to 3.5 $\mu$m.
Our careful data processing suggests that most of the DIRBE polarimetric observations are low signal-to-noise ratio (SNR) data, and that this is the reason these data have not been analyzed for many years.
Therefore, we carefully extracted only high SNR data, and we use the data around $\epsilon$ = 90$^{\circ}$ to study the characteristics of ZL polarization.

\section{Methodology}
\subsection{DIRBE Weekly Sky Maps} \label{sec:DIRBE}
\begin{figure*}[tbp]
\epsscale{1.2}
\plotone{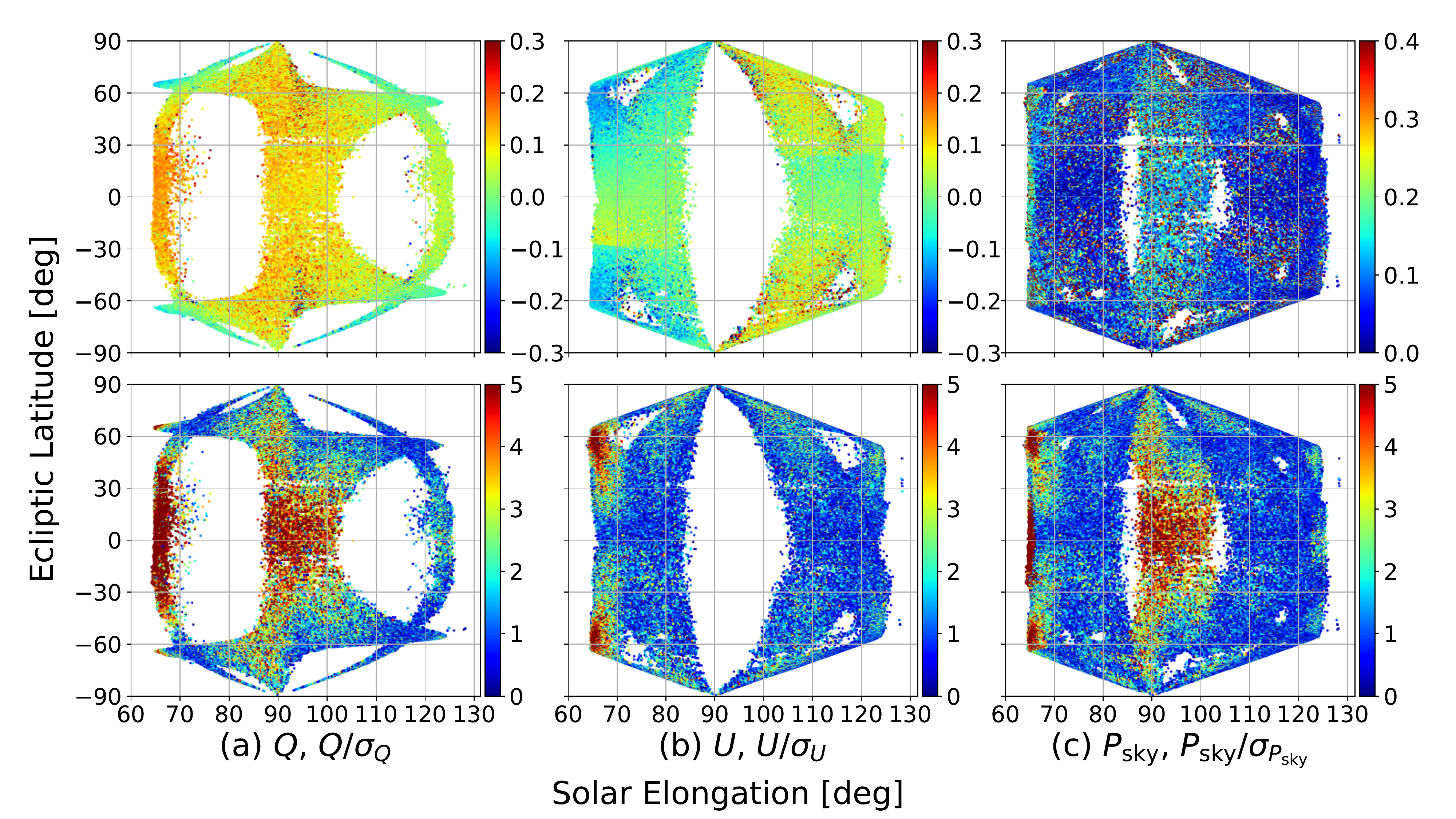}
\caption{Top: sky maps of Stokes parameters (a) $Q$, (b) $U$ and (c) degree of polarization of the sky brightness $P_{\rm sky}$ at 1.25 $\mu$m from 1989 December 11 to December 17.
Bottom: SNR maps (a: $Q/\sigma_{Q}$, b: $U/\sigma_{U}$, c: $P_{\rm sky}/\sigma_{P_{\rm sky}}$) with the same data as top.
}
\label{fig:QUPmap}
\end{figure*}

DIRBE is one of three instruments on board the COBE satellite \citep{1991AIPC..222..161H,1992ApJ...397..420B,1993SPIE.2019..180S}.
In addition to photometric observations, DIRBE was designed to make measurements of linear polarization in three bands centered at 1.25, 2.2, and 3.5 $\mu$m.
The DIRBE optical axis is oriented at 30$^{\circ}$ from the COBE spacecraft spin axis and observed half of the sky each day at solar elongation angles ranging from $\epsilon$ = 64$^{\circ}$ to 124$^{\circ}$.
The Weekly Sky Maps \footnote{\href{https://lambda.gsfc.nasa.gov/product/cobe/dirbe_cwm_data_get.html}{https://lambda.gsfc.nasa.gov/product/cobe/dirbe\_cwm\_data\_get.html}} consist of 41 files, one per week of optimized operation of the cryogenic mission from November 24, 1989 to September 21, 1990 \citep{hauser1998cobe}.
For telemetry data from DIRBE, data reduction and absolute calibration of the available maps have already been done by the DIRBE team. 
The Weekly Sky Maps give Stokes $Q$ and $U$ parameters at 1.25, 2.2, and 3.5 $\mu$m.
The Stokes parameters for each band are averages of the individual $Q$ and $U$ observations, weighted by their measurement uncertainties.
The standard deviations of the weighted mean values of the Stokes parameters are $\sigma_{Q}$ and $\sigma_{U}$, respectively.
$Q$ is the component parallel (positive) or perpendicular (negative) to the local line of ecliptic longitude, and $U$ is the component 45$^{\circ}$ E (positive) or 45$^{\circ}$ W (negative) to the local line of ecliptic longitude.
The degree of polarization $P$ and its standard deviation $\sigma_{P}$ are calculated using $Q$, $U$, $\sigma_{Q}$, and $\sigma_{U}$ as 
\begin{equation}
P=\left\{
\begin{array}{cl}
Q\cos{2\theta}+U\sin{2\theta}, & \text{if both $Q$ and $U$ exist} \\
Q/\cos{2\theta}, & \text{if only $Q$ exists} \\
U/\sin{2\theta}, & \text{if only $U$ exists} 
\end{array}
\right.
\label{eq:pol}
\end{equation}
\begin{equation}
\sigma_{P}=\left\{
\begin{array}{cl}
P\sqrt{{\bigl( \frac{\sigma_{Q}}{Q} \bigl)}^2+{\bigl( \frac{\sigma_{U}}{U} \bigl)}^2}, & \text{if both $Q$ and $U$ exist} \\
P\frac{\sigma_{Q}}{Q}, & \text{if only $Q$ exists} \\
P\frac{\sigma_{U}}{U}, & \text{if only $U$ exists} \\
\end{array}
\right.
\label{eq:pol2}
\end{equation}
where $\theta$ is the orientation of the polarizer, defined as the angle between the polarizer axis and the local meridian.
$\theta$ is computed in the range $[-90^{\circ}, \, +90^{\circ}]$.
Since $P$ is always a positive quantity unlike $Q$ and $U$, which can be positive or negative, the naive estimate $P = \sqrt{Q^2+U^2}$ is biased for low SNR data.
$\sigma_{P}$ tends to be larger when $P$ is evaluated from both $Q$ and $U$ than when $P$ is evaluated using only $Q$ because of the large $\sigma_{U}$ in the data. However, the difference in $P$ is less than 10\%.
The mean observation time and mean solar elongation angle for each sky region are also recorded in Weekly Sky Maps.
Further information about data processing of the DIRBE polarization can be found in \citet{hauser1998cobe}.

\begin{figure*}[tbp]
\epsscale{1.2}
\plotone{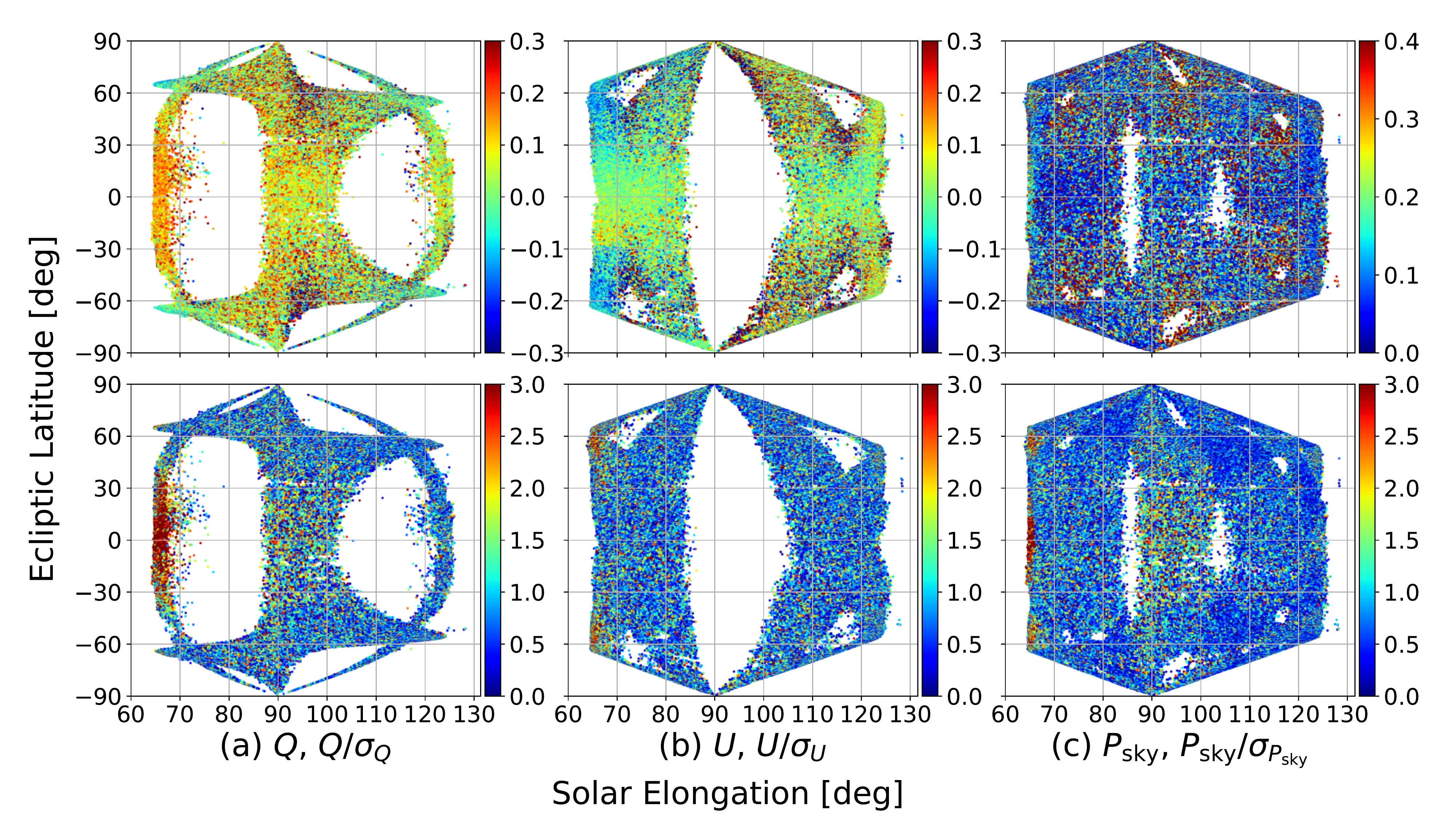}
\caption{Same as Figure \ref{fig:QUPmap}, but at 2.2 $\mu$m from 1989 December 11 to December 17.
}
\label{fig:QUPmap2}
\end{figure*}
\begin{figure*}[tbp]
\epsscale{1.2}
\plotone{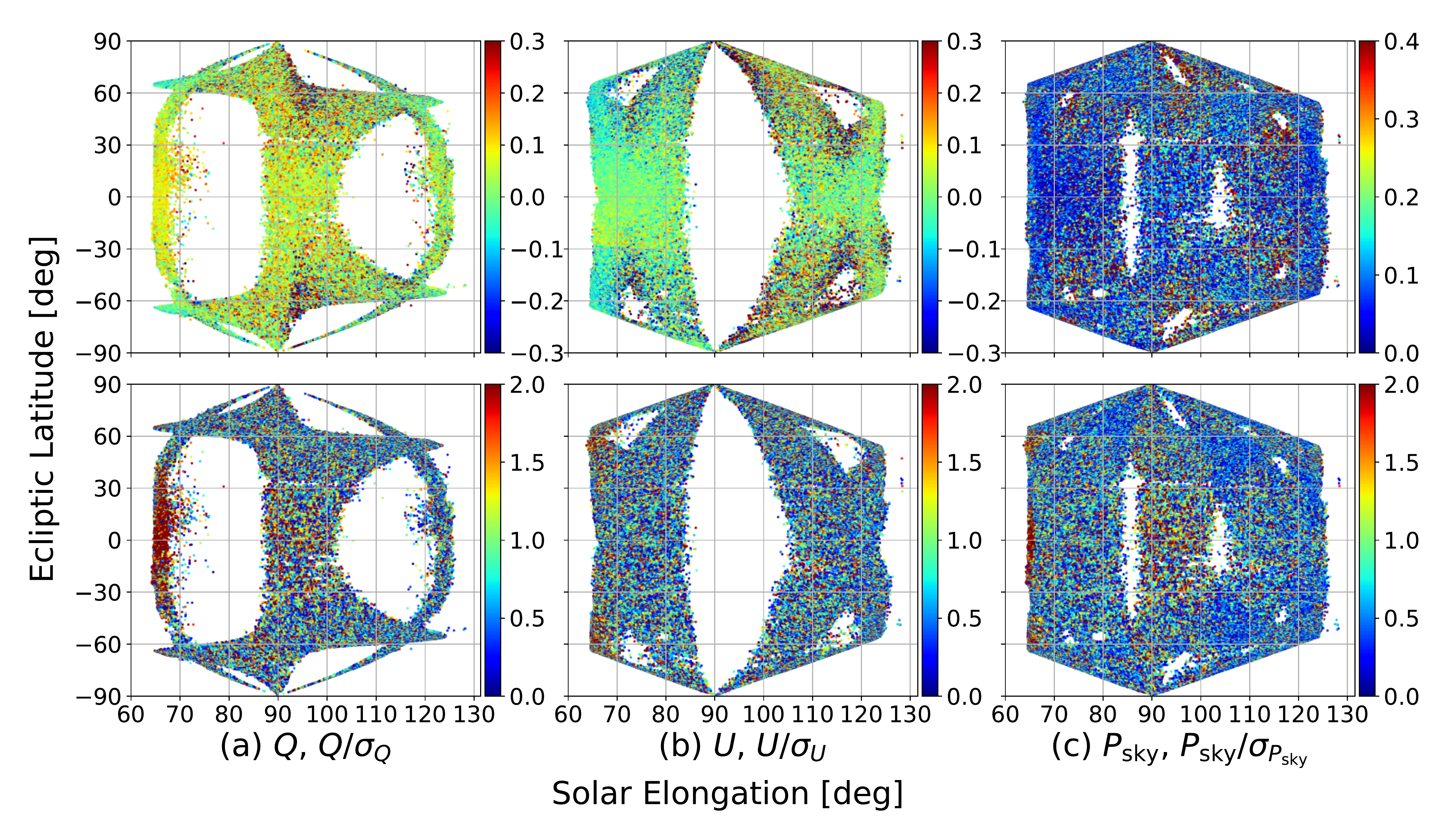}
\caption{Same as Figure \ref{fig:QUPmap}, but at 3.5 $\mu$m from 1989 December 11 to December 17.
}
\label{fig:QUPmap3}
\end{figure*}


All 41 Weekly Sky Map files have file extensions in FITS format.
A quadrilateralized spherical projection and a quad–tree pixelization scheme were adopted for all COBE sky maps, including DIRBE maps. 
The COBE Quadrilateralized Spherical Cube (CSC) is an approximately equal–area projection in which the celestial sphere is projected onto an inscribed cube.
The coordinate system is Geocentric Ecliptic J2000. 
An ancillary information file called {\tt DIRBE\_SKYMAP\_INFO.FITS} explains the quad–tree scheme that relates CSC pixel numbers to ($x$, $y$) positions within the quad–sphere cube faces and tabulates the ecliptic J2000, Galactic ($l$, $b$) and equatorial J2000 coordinates corresponding to the DIRBE pixel centers. 
This file may be obtained from the DIRBE ancillary data products page\footnote{\href{https://lambda.gsfc.nasa.gov/product/cobe/dirbe_ancil_qs_get.html}{https://lambda.gsfc.nasa.gov/product/cobe/dirbe\_ancil\_qs\_get.html}}.

For applying data cuts and checking for systematics and dependencies, we convert to several coordinate systems.
After converting the maps to ecliptic, Galactic, and equatorial coordinates, the following processing is applied to each map.
Only data in the high Galactic latitude region ($|b|>30^{\circ}$) are extracted, since data in the Galactic plane are not suitable for studying ZL polarization.
In addition, we exclude data obtained when DIRBE line of sight approached certain solar system objects to avoid contaminating the data by light from moving objects.
We also remove noisy data in regions with fewer than 10 observations per week, or when the calculation of Stokes parameters by robust fitting part of DIRBE original data processing fails, or when the processing of polarization ratios fails to produce $Q$ or $U$ values.
The degree of polarization derived by Equation (\ref{eq:pol}) using the raw DIRBE data is the ratio of the intensity of the total sky to the intensity of the polarization component.
The polarization of the sky brightness, $P=P_{\rm sky}$, can be written as 
\begin{equation}
P_{\mathrm{sky}}= \frac{I_{\mathrm{pol}}}{I_{\mathrm{sky}}},
\end{equation}
where $I_{\mathrm{pol}}$ is the intensity of polarized light component from the DIRBE polarization channel and $I_{\mathrm{sky}}$ is the intensity obtained from the DIRBE intensity channel.

Figure \ref{fig:QUPmap}-\ref{fig:QUPmap3} represent sky maps of Stokes parameters and degree of polarization of the sky brightness from 1.25 to 3.5 $\mu$m for a week.
$P_{\rm sky}$ map at 1.25 $\mu$m for solar elongation below 70$^{\circ}$ or near 90$^{\circ}$ is more reliable than other regions because a SNR ($P_{\rm sky}/\sigma_{P_{\rm sky}}$) is more than three.
Uncertainties on the polarization depend on the coverage map in both $Q$ and $U$. 
Since ZL brightness decreases toward longer wavelengths in the near-infrared, same as the solar spectrum, the SNRs of the observed data at 2.2 and 3.5 $\mu$m are smaller than those at 1.25 $\mu$m.
The total number of $P_{\rm sky}$ data calculated for all weekly data is 1,624,852 at 1.25 $\mu$m, 2,645,893 at 2.2 $\mu$m, and 2,645,681 at 3.5 $\mu$m, respectively.
The difference in the total number is due to unexplained malfunctions in each polarization channel, especially in "b" channel, where 1.25 $\mu$m is inoperative for about 35\% of the observation period. 
This channel is especially noisy for observations in which the Moon is visible in the viewing swath. 
\begin{figure}[tbp]
\epsscale{1.2}
\plotone{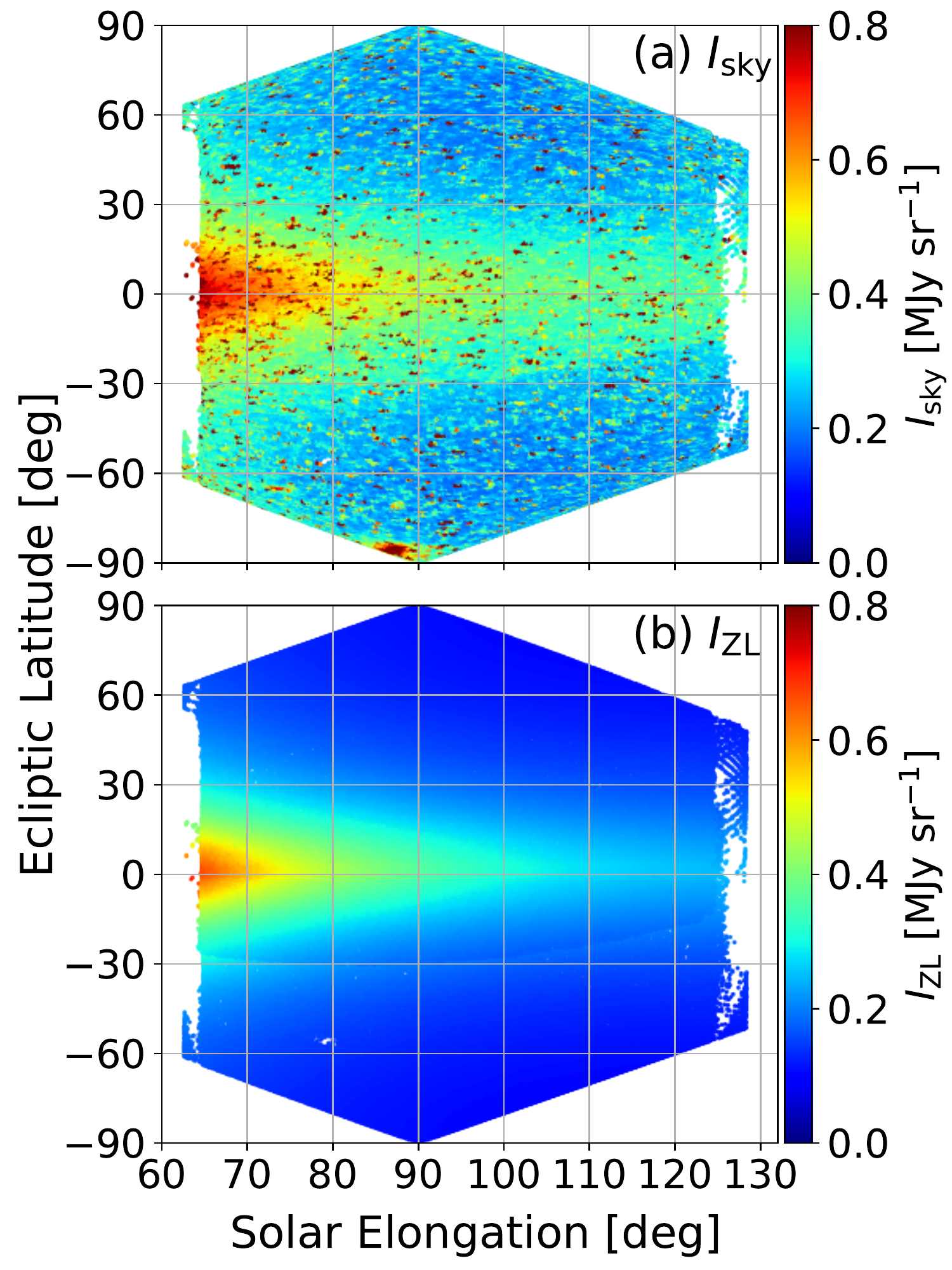}
\caption{(a) Intensity map obtained from the DIRBE intensity channel at 1.25 $\mu$m from 1989 December 11 to December 17. 
(b) ZL intensity map estimated by DIRBE IPD model \citep{1998ApJ...508...44K} at 1.25 $\mu$m.}
\label{fig:se_beta_week04_I1}
\end{figure}
\begin{figure*}[tbp]
\epsscale{1.2}
\plotone{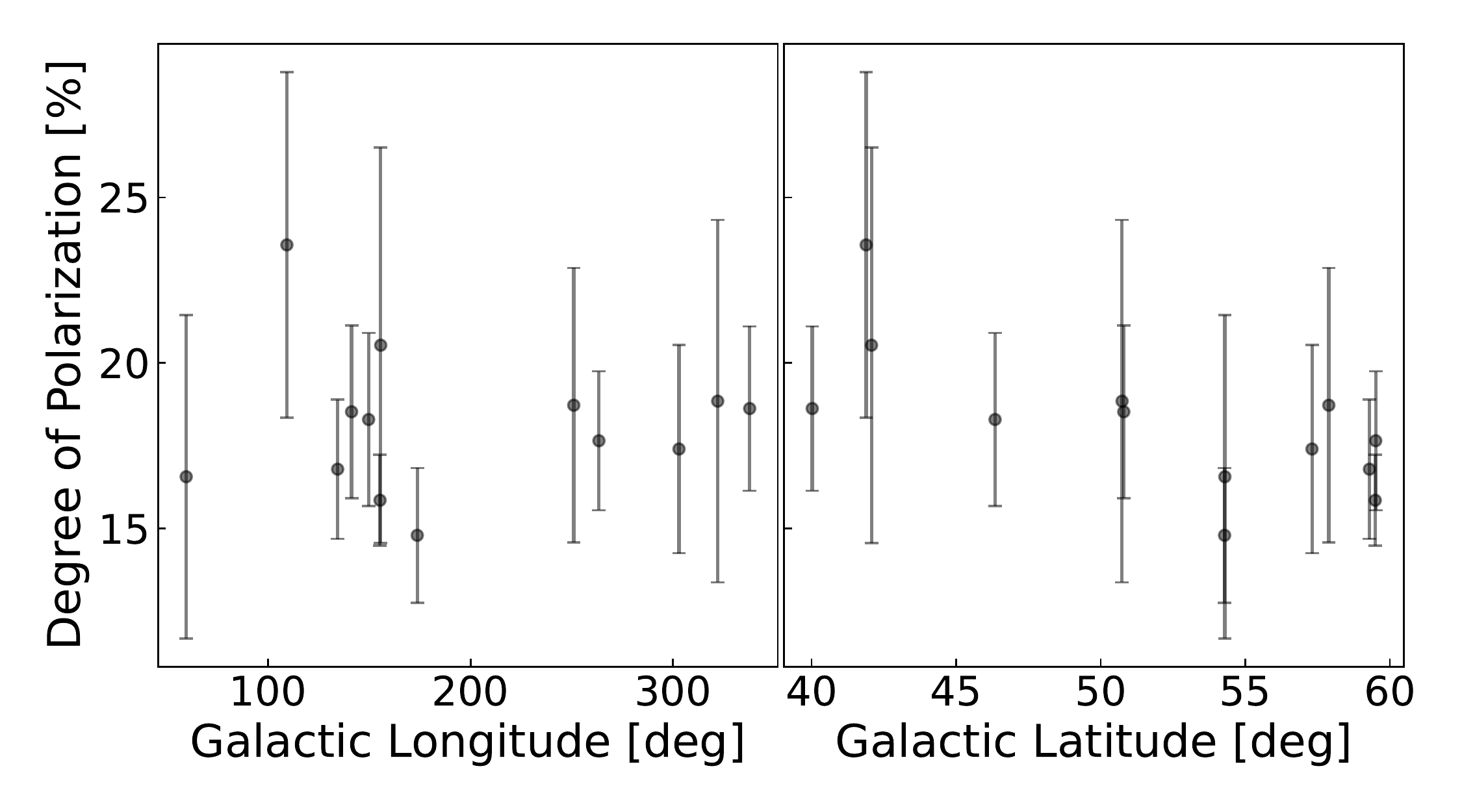}
\caption{The Galactic coordinate dependence of $P_{\rm ZL}$ at $\lambda$ = 1.25 $\mu$m and ($\epsilon$, $\beta$) = (90$^{\circ}$, 0$^{\circ}$).
Plots are weekly results, smoothed by a weighted average of the region averaged over one square degree.
}
\label{fig:WeekConsistency}
\end{figure*}
\begin{figure*}[tbp]
\epsscale{1.2}
\plotone{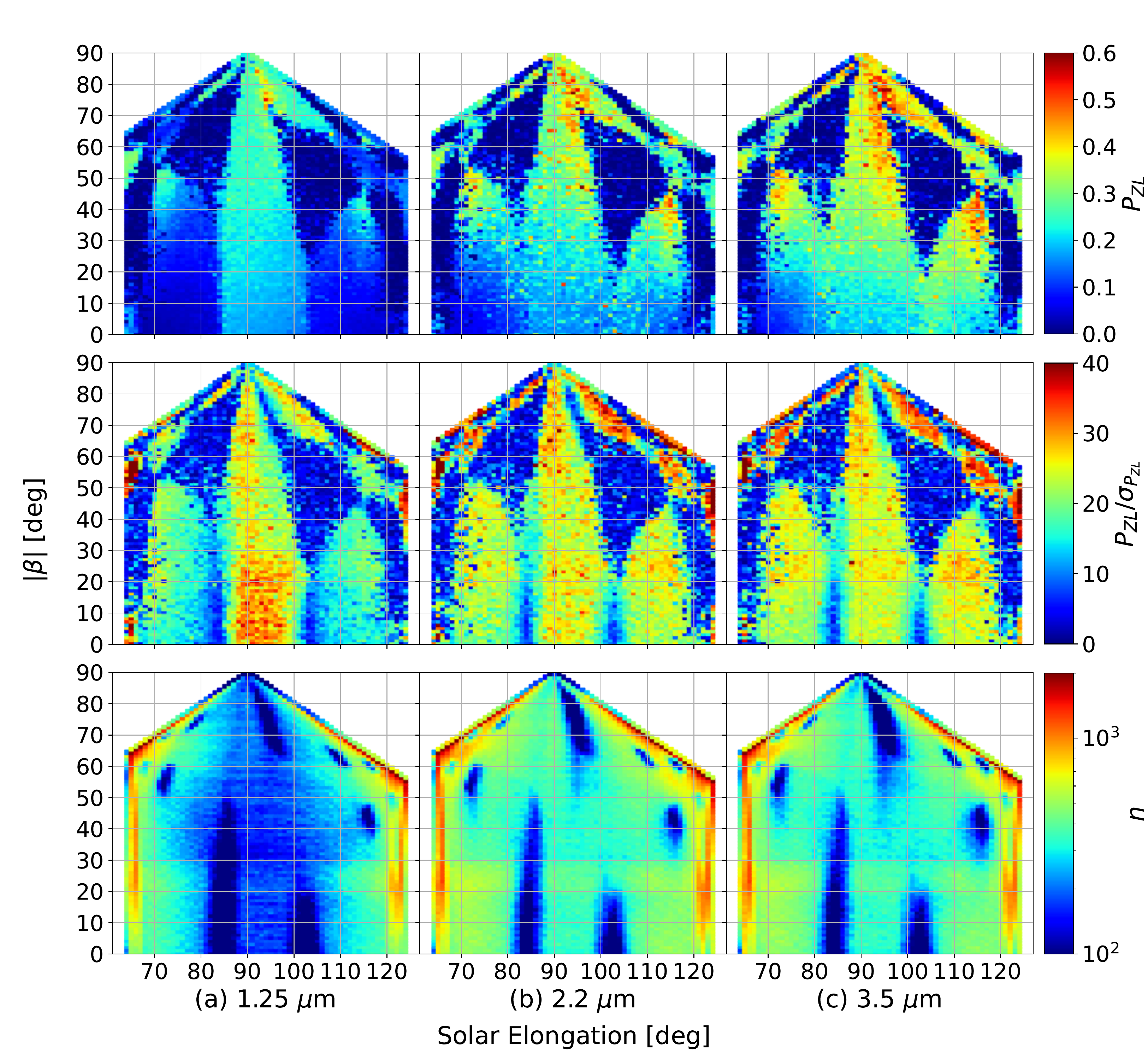}
\caption{Top: zodiacal light polarization map for each wavelength(a: 1.25 $\mu$m, b: 2.2 $\mu$m, c: 3.5 $\mu$m) calculated by weighted average per one square degree.
Middle: SNR map.
Bottom: number of zodiacal light polarization data map.
}
\label{fig:Pmeanmap}
\end{figure*}

\subsection{Polarization Measurement}
From measurements of polarization on the sky and surface brightness measurements, we derive estimates of the ZL polarization, $P_{\rm ZL}$.
We assume the polarization components of the other diffuse radiation are negligible compared to ZL. 
$I_{\mathrm{sky}}$ denotes the integrated surface brightness from ZL, $I_{\mathrm{ZL}}$, along with contributions from bright point sources and other diffuse radiation, including diffuse galactic light (DGL), integrated starlight (ISL) and the extragalactic background light (EBL).
Bright galaxies are also a source of astrophysical noise for the DIRBE measurement.
In addition, $I_{\mathrm{sky}}$ at 3.5 $\mu$m is contaminated by thermal emission component of the ZL from IPD particles.
The corrected polarization, $P_{\rm ZL}$, is expressed as:
\begin{equation}
P_{\mathrm{ZL}}=P_{\mathrm{sky}} \frac{I_{\mathrm{sky}}}{I_{\mathrm{ZL}}},
\end{equation}
where $I_{\mathrm{ZL}}$ is the ZL intensity, which is only sunlight scattering, at each wavelength derived in Equation (\ref{eq:kel}) \citep{1998ApJ...508...44K}.
Figure \ref{fig:se_beta_week04_I1} shows sky maps of $I_{\mathrm{sky}}$ and $I_{\mathrm{ZL}}$ at 1.25 $\mu$m for a week.
The $I_{\mathrm{sky}}$ map shows the contribution of stars and the Large Magellanic Cloud (($\epsilon$, $\beta$) $\sim$ (88$^{\circ}$, -85$^{\circ}$)). 
$I_{\mathrm{sky}}$ is also higher than $I_{\mathrm{ZL}}$ in all regions due to other diffuse light.
We check for temporal variation in the polarization before averaging results across all weeks.
Figure \ref{fig:WeekConsistency} shows the Galactic coordinate dependence of $P_{\rm ZL}$ around $\epsilon=90^{\circ}$ at $\beta=0^{\circ}$.
The standard deviation across weeks in that region is 2.11\%. The combined data are consistent up to the measurement uncertainty.
Figure \ref{fig:Pmeanmap} shows $P_{\rm ZL}$ map, $P_{\rm ZL}/\sigma_{P_{\rm ZL}}$ map, and number of $P_{\rm ZL}$ data, $n$ map at each wavelength.
Each map is smoothed by a weighted mean of regions over one square degree, with no north-south distinction of ecliptic latitude, by using data from all weekly observations.
$\sigma_{P_{\mathrm{ZL}}}$ is the standard deviation of the weighted mean of the degree of polarization.
Around $\epsilon=90^{\circ}$, $P_{\rm ZL}$ map at 1.25 $\mu$m shows high SNR more than twenty five despite the small $n$.
Even though $\sigma_{P_{\mathrm{ZL}}}$ is inversely proportional to the square root of $n$, the high SNR confirms that this region is more reliable than other regions.As already mentioned in Section \ref{sec:DIRBE}, the standard deviations, such as $\sigma_{Q}$, $\sigma_{U}$, and $\sigma_{P_{\rm sky}}$, near the region where neither the Stokes parameter $Q$ nor $U$ exists is larger than those in other regions. 
Therefore, even if $n$ is large and $P_{\rm ZL}/\sigma_{P_{\rm ZL}}$ is high, the reliability is low in certain regions (e.g., ($\epsilon$, $\beta$) = (115$^{\circ}$, $\pm$30$^{\circ}$), (95$^{\circ}$, $\pm$60$^{\circ}$), (75$^{\circ}$, $\pm$45$^{\circ}$)).
For these reasons, we use only the region around $\epsilon=90^{\circ}$ to study the $\epsilon$-, $\beta$-, and $\lambda$-dependence.

\section{Results and Discussions} \label{sec:result}

\subsection{Solar Elongation Dependence}

\begin{figure}[htbp]
\epsscale{1.2}
\plotone{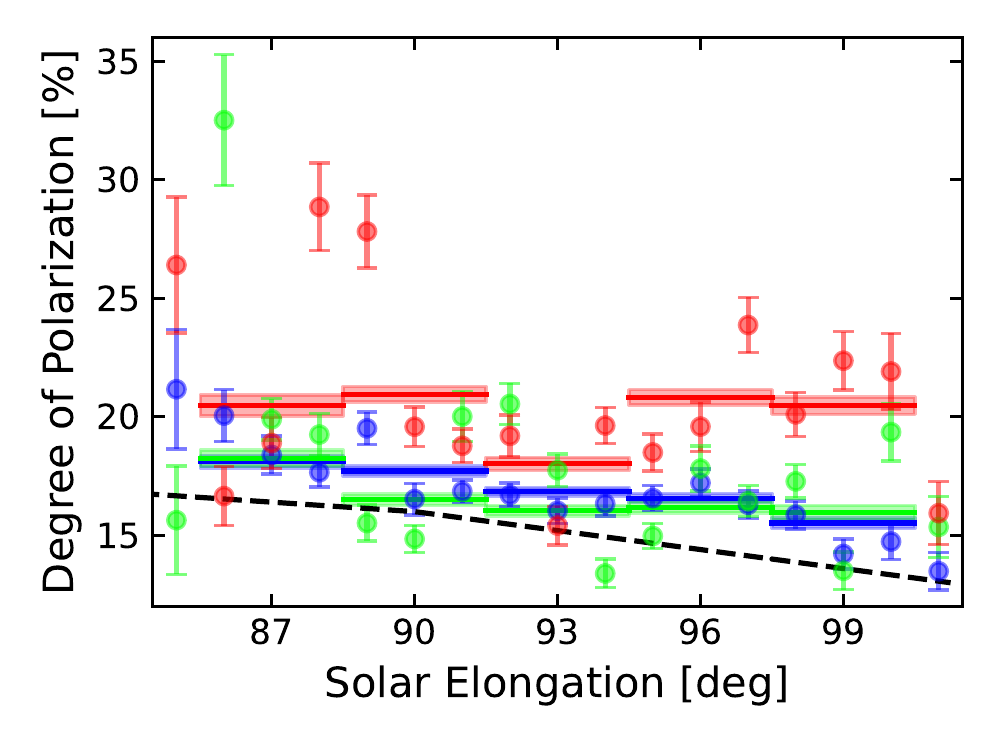}
\caption{The solar elongation dependence of $P_{\rm ZL}$ at $\beta=0^{\circ}$.
The filled circles are the result of this work, smoothed by a weighted average of the region averaged over one square degree (blue: 1.25 $\mu$m, green: 2.2 $\mu$m, red: 3.5 $\mu$m).
The solid lines are the result of this work, smoothed by a weighted mean of regions over three square degrees.
The error bars and the shaded regions indicate the standard deviation of the weighted mean of the degree of polarization.
The dashed line represents the ZL polarization at $\lambda$ = 550 nm \citep{1996ASPC..104..301L,1998A&AS..127....1L}.
}
\label{fig:plot_P1_se_paper}
\end{figure}

Since the Stokes parameter $U$ is zero in the ecliptic plane, $\epsilon$-dependence of $P_{\rm ZL}$ is valid only in the measurement region of the Stokes parameter $Q$.
Figure \ref{fig:plot_P1_se_paper} shows the $\epsilon$-dependence of $P_{\rm ZL}$ around $\epsilon=90^{\circ}$ at $\beta=0^{\circ}$.
$P_{\rm ZL}$ near the ecliptic plane shows little wavelength dependence, differing only by a few percent.
At $\lambda$ = 1.25 $\mu$m, $P_{\rm ZL}$ tends to decrease as $\epsilon$ increases, which can be explained by the fact that the degree of polarization generally decreases with increasing scattering angle above $90^{\circ}$ \citep{1978A&A....65..265G}.
This trend is also consistent with the degree of polarization observed from Earth at $\lambda$ = 550 nm \citep{1996ASPC..104..301L,1998A&AS..127....1L}.
On the other hand, there is no significant $\epsilon$-dependence of $P_{\rm ZL}$ at 2.2 and 3.5 $\mu$m, due to large standard deviation of the data. 
Another factor may be that the scattering angle dependence of degree of polarization of IPD is different in the near-infrared.
As the wavelength increases, $P_{\rm ZL}$ increases, and the $\epsilon$-dependence becomes weak not because of real de-correlation, but as a result of noisy observations.

\subsection{Ecliptic Latitude Dependence}

\begin{figure}[htbp]
\epsscale{1.2}
\plotone{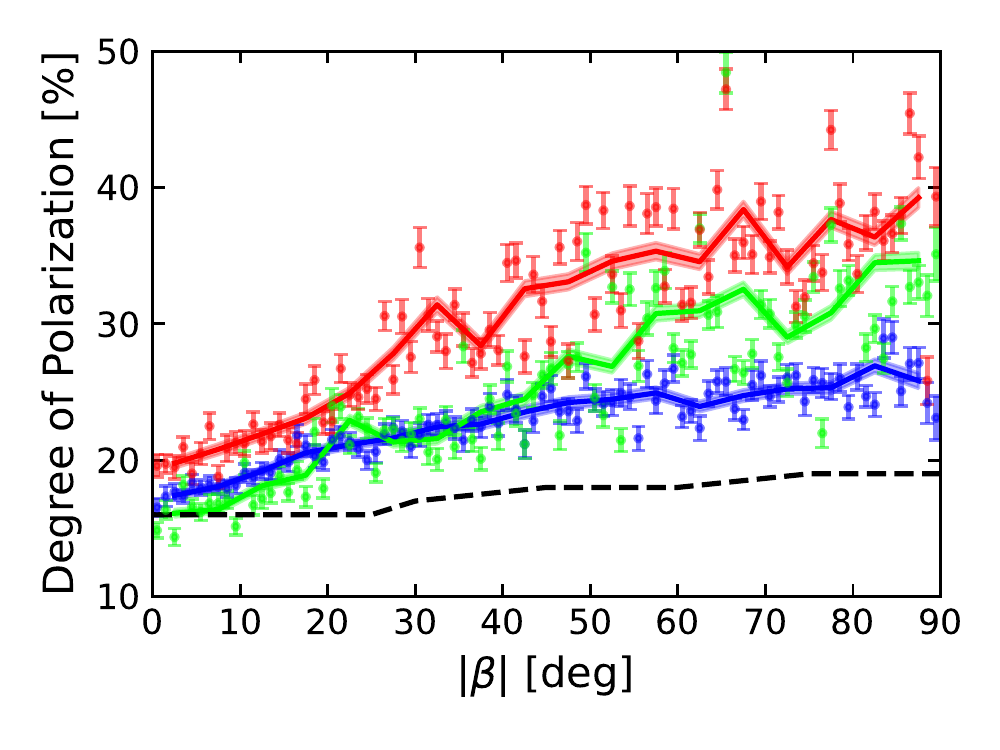}
\caption{The ecliptic latitude dependence of $P_{\rm ZL}$ at $\epsilon=90^{\circ}$.
The filled circles are the result of this work, smoothed by a weighted average of the region averaged over one square degree (blue: 1.25 $\mu$m, green: 2.2 $\mu$m, red: 3.5 $\mu$m).
The solid lines are the result of this work, smoothed by a weighted mean of regions averaged over five square degrees.
The error bars and the shaded regions indicate the standard deviation of the weighted mean of the degree of polarization.
The dashed line represents the ZL polarization at $\lambda$ = 550 nm \citep{1996ASPC..104..301L,1998A&AS..127....1L}.
}
\label{fig:plot_P1_el_paper}
\end{figure}

We compare the degree of polarization from the ecliptic plane to the ecliptic pole at $\epsilon=90^{\circ}$.
$P_{\rm ZL}$ as a function of the ecliptic latitude $\beta$ at $\epsilon=90^{\circ}$ is shown in Figure \ref{fig:plot_P1_el_paper}.
At all observed wavelengths, $P_{\rm ZL}$ tends to increase as $|\beta|$ increases.
This trend is also consistent with $P_{\rm ZL}$ at $\lambda$ = 550 nm.
This $\beta$-dependence can be explained by the fact that as $|\beta|$ increases, there are less IPDs farther away in the line-of-sight direction, and the majority of $P_{\rm ZL}$ is produced by IPDs near the earth.
The $\beta$-dependence is larger at longer wavelengths with $P_{\rm ZL}$ at 3.5 $\mu$m about twice as large as that at 550 nm in the ecliptic pole region.

\subsection{Wavelength Dependence}
\begin{figure}[tbp]
\epsscale{1.2}
\plotone{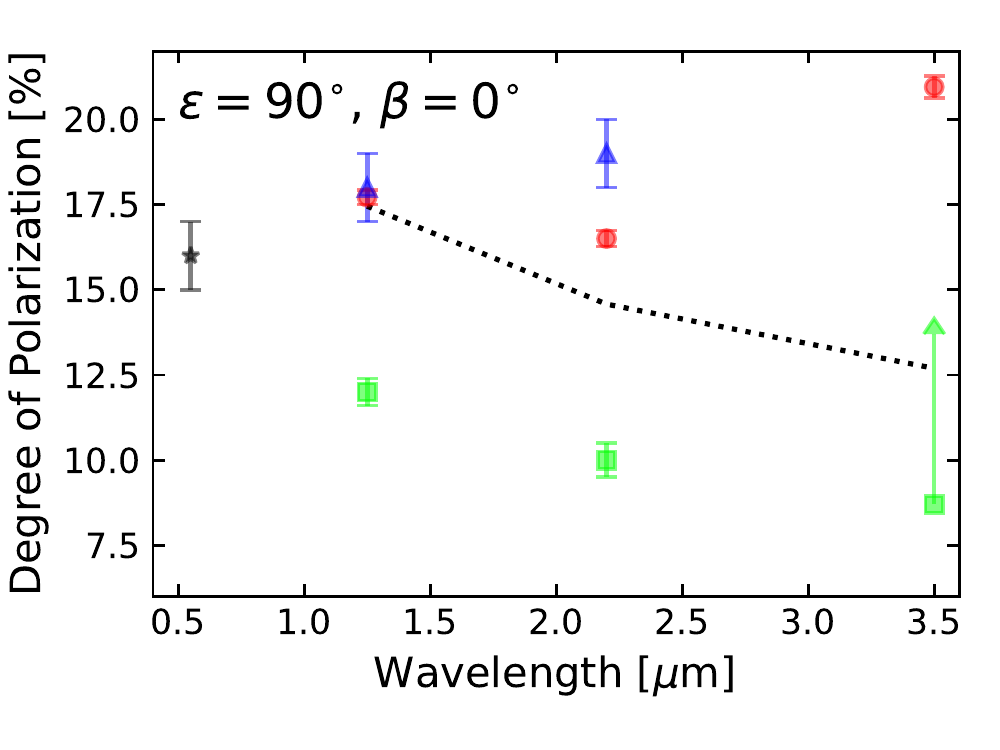}
\caption{The wavelength dependence of $P_{\rm ZL}$ at $\epsilon=90^{\circ}$ and $\beta=0^{\circ}$.
The red circles are the results of this work.
The error bars indicate the standard deviation of the weighted mean of the degree of polarization.
The degree of polarization calculated by \citet{1994ApJ...431L..63B} is shown as green squares (1.25 and 2.2 $\mu$m) and arrow (lower limit at 3.5 $\mu$m).
The blue triangles are the results calculated by \citet{Takimoto_2022}.
The black star indicates the zodiacal light polarization at 0.55 $\mu$m \citep{1996ASPC..104..301L}.
The dotted line represent the degree of polarization calculated by the geometrical-optical scattering with graphite \citep{1984ApJ...285...89D}.
}
\label{fig:plot_P1_wl_paper}
\end{figure}

Figure \ref{fig:plot_P1_wl_paper} shows $P_{\rm ZL}$ as a function of the wavelength $\lambda$ at $\epsilon=90^{\circ}$ and $\beta=0^{\circ}$.
$P_{\rm ZL}$ at 1.25, 2.2, and 3.5 $\mu$m are 17.7 $\pm$ 0.2\%, 16.5 $\pm$ 0.2\%, and 21.0 $\pm$ 0.3\%, respectively.
Our results show that there is little wavelength dependence of $P_{\rm ZL}$ near the ecliptic plane from visible to near-infrared.
Compared to the average value from \citet{1994ApJ...431L..63B}, our results are more than 5\% greater.
On the other hand, the degree of polarization at 1.25 and 2.2 $\mu$m removed starlight and other contributions by \citet{Takimoto_2022} are consistent with this work.
The polarization due to graphite \citep{1984ApJ...285...89D} with particle radii greater than 10 $\mu$m, based on Mie theory calculations, shows a tendency to decrease toward longer wavelengths, different from our results.

\begin{figure}[tbp]
\epsscale{1.2}
\plotone{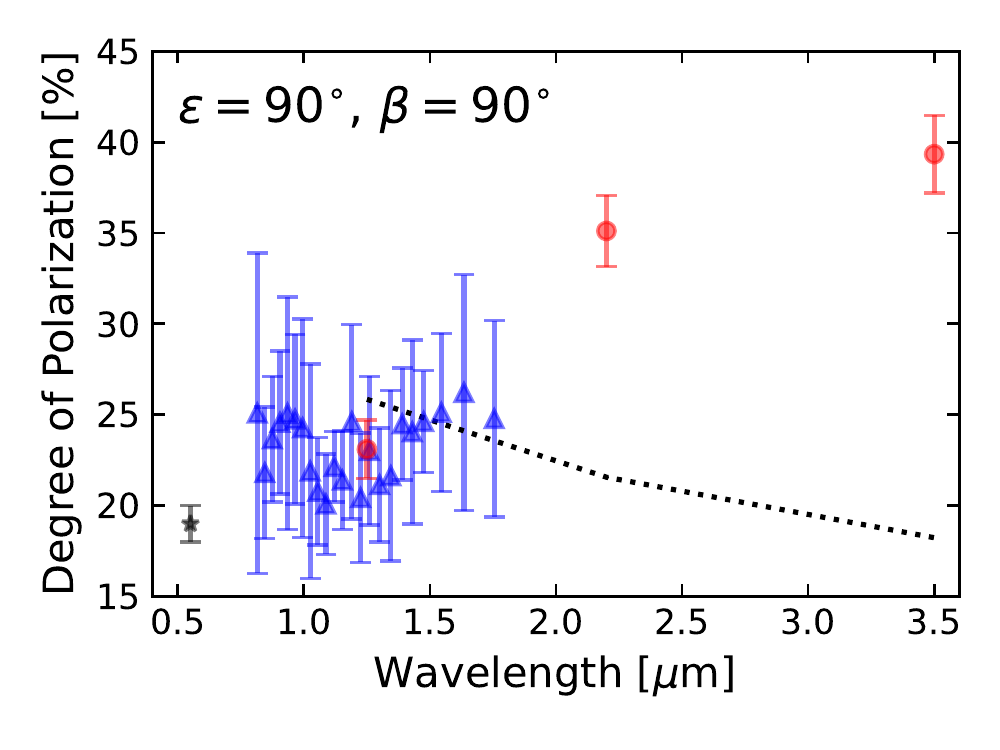}
\caption{The wavelength dependence of $P_{\rm ZL}$ at the NEP field.
The red circles are the results of this work.
The error bars represent the standard deviation of the weighted mean of the degree of polarization.
The blue triangles are the zodiacal light polarization spectrum measured with CIBER \citep{Takimoto_2022}.
The black star indicates the zodiacal light polarization at 0.55 $\mu$m \citep{1996ASPC..104..301L}.
The dotted line represents the degree of polarization calculated by Mie scattering using graphite with a particle radius of 10 $\mu$m \citep{1984ApJ...285...89D}.
}
\label{fig:plot_P1_wl90_paper}
\end{figure}

$P_{\rm ZL}$ as a function of the wavelength $\lambda$ in the NEP field is shown in Figure \ref{fig:plot_P1_wl90_paper}.
Our result at $\lambda$ = 1.25 $\mu$m is $P_{\rm ZL}$ = 23.1 $\pm$ 1.6\%, consistent with CIBER results.
In addition, our works at $\lambda$ = 2.2 and 3.5 $\mu$m show $P_{\rm ZL}$ = 35.1 $\pm$ 2.0\% and 39.3 $\pm$ 2.1\%, respectively.
Our results suggest that $P_{\rm ZL}$ tends to increase toward longer wavelengths in the visible to near-infrared.
From CIBER observations, the polarization properties of ZL at 1.25 $\mu$m can be explained if the IPD particles are dominated by absorptive materials with a particle radius of 1 $\mu$m or larger \citep{Takimoto_2022}. 
On the other hand, a candidate IPD particle that provides a high degree of polarization at wavelengths longer than 2.2 $\mu$m requires a size equal to or smaller than the wavelength. 
Absorptive particles larger than 10 $\mu$m diameter exhibit geometrical-optical scattering characteristics.
Thus, when particles larger than 10 $\mu$m dominate, the ZL polarization indicates a decreasing trend toward longer wavelengths, so the results measured by DIRBE cannot be reproduced.
Therefore, it is suggested that absorptive particles with a radius between 1 $\mu$m and 10 $\mu$m dominate the polarization properties.
Another candidate is porous silicate grains with a power-law number distribution by mass of $dn/d\log m \propto m^{\alpha}$, where $\alpha = -0.56$ \citep{1994ApJ...431L..63B}.
It is similar to the model that \citet{1994ApJ...432L..71L} found to be applicable to comets.
Therefore, it is suggested that the IPD drifting in the high ecliptic latitude region is cometary dust composed of porous silicates.
Table \ref{tbl:pol} summarizes the observation fields and degree of polarization of ZL.

\begin{deluxetable}{cccc}
\label{tbl:pol}
\tabletypesize{\scriptsize}
\tablecaption{Near-infrared Polarization of The Zodiacal Light Observed with DIRBE}
\tablewidth{0pt}
\tablehead{\colhead{Wavelength} & \colhead{Ecliptic Lat.}&\colhead{Solar Elong.} & \colhead{Degree of Pol.} \\
\colhead{$\lambda$} & \colhead{$\beta$}&\colhead{$\epsilon$} & \colhead{$P_{\rm ZL}$} \\
\colhead{($\mu$m)}& \colhead{(deg)} &\colhead{(deg)} & \colhead{(\%)}}
\startdata
1.25 & 0 & 90 & 17.7 $\pm$ 0.2\\
& 90 & 90 & 23.1 $\pm$ 1.6\\
2.2 & 0 & 90 & 16.5 $\pm$ 0.2\\
& 90 & 90 & 35.1 $\pm$ 2.0\\
3.5 & 0 & 90 & 21.0 $\pm$ 0.3\\
& 90 & 90 & 39.3 $\pm$ 2.1\\
\enddata
\end{deluxetable}
The polarization properties of ZL obtained with DIRBE can be a crucial tool for constraining the various properties of IPD. 
However, we cannot determine the origin of IPDs from our data alone. 
Future work is necessary to better constrain the properties of the IPD particles. 
We need to devise a new IPD spatial distribution model that includes an isotropic component of ZL \citep{2020ApJ...901..112S}.
The polarization properties of IPD particles must be investigated in more detail.
More theoretical modeling of IPD scattering is needed to determine the origin of IPD.
Future model simulations will need to carefully consider the wavelength-specific scattering properties of the constituent minerals to improve our understanding of the structure and physical properties of IPD.
In addition, incorporating details of the complex particle shape and size distribution into the model should make it possible to reproduce the observed ZL polarization.
Further polarization data at solar elongation, which were not observed in DIRBE, are needed to limit the composition of IPD. 
Forward scattering ($\Theta<90^{\circ}$) shows characteristic composition-dependent scattering properties and is effective for validating compositional models of IPDs.
Therefore, near-infrared ZL polarization observations at $\epsilon<60^{\circ}$ are especially necessary to capture forward scattering components.

\section{Conclusion} \label{sec:summary}
We described the near-infrared polarization of the ZL measured from space by DIRBE/COBE in the discrete photometric bands of 1.25, 2.2, and 3.5 $\mu$m.
To make ZL polarimetric maps using the all data from the DIRBE Weekly Sky Maps, we replaced the sky intensity, which includes contributions from starlight and other sources from observations, with a ZL intensity model.
These maps show the characteristics of different SNRs for each sky region and wavelength of observation.
New analysis in terms of solar elongation, ecliptic latitude, and wavelength suggest that the ZL polarization of the near-infrared can be explained by an absorptive particle model with a few microns size.
For the 1.25 $\mu$m ZL polarization map, the degree of polarization of the NEP region is consistent with CIBER observations, and that near the ecliptic plane is also comparable to DIRBE observations, which were estimated using different methods.
More detailed investigation of the particle properties is needed in the future.

COBE is supported by NASA's Astrophysics Division.
Goddard Space Flight Center (GSFC), under the scientific guidance of the COBE Science Working Group, is responsible for the development and operation of COBE.

\bibliography{sample631}{}
\bibliographystyle{aasjournal}



\end{document}